\documentclass[final,5p,times,twocolumn,nopreprintline]{elsarticle}
\usepackage{amsmath,slashed,booktabs}
\usepackage{graphicx,graphics}
\usepackage{dcolumn}
\usepackage[hyperfootnotes=false]{hyperref}
\usepackage{xspace}
\usepackage{color}
\usepackage{balance}
\usepackage{multirow}
\usepackage{multicol}

\usepackage{fancyhdr}
\fancypagestyle{firstpage}{%
	
	\lhead{}
	\rhead{\small INT-PUB-22-024, PSI-PR-22-28, ZU-TH 43/22}
}
\newcommand{\GeV}{\,\text{GeV}}
\newcommand{\MeV}{\,\text{MeV}}

\newcommand{\mpi}{M_\pi}
\newcommand{\mk}{M_K}
\newcommand{\eps}{\epsilon}

\newcommand{\Order}{\mathcal{O}}
\newcommand{\beq}{\begin{equation}}
\newcommand{\eeq}{\end{equation}}

\allowdisplaybreaks[4]

\begin{document}

\renewcommand{\theequation}{\arabic{equation}}

\begin{frontmatter}
 
\title{Scrutinizing CKM unitarity with a new measurement of the $K_{\mu3}/K_{\mu 2}$ branching fraction}

\author[Seattle]{Vincenzo Cirigliano}
\author[PSI,Zurich]{Andreas Crivellin}
\author[Bern]{Martin Hoferichter}
\author[Frascati]{Matthew Moulson}

\address[Seattle]{Institute for Nuclear Theory, University of Washington, Seattle WA 91195-1550, USA}
\address[PSI]{Paul Scherrer Institut, 5232 Villigen PSI, Switzerland}
\address[Zurich]{Physik-Institut, Universit\"at Z\"urich, Winterthurerstrasse 190, 8057 Z\"urich, Switzerland}
\address[Bern]{Albert Einstein Center for Fundamental Physics, Institute for Theoretical Physics, University of Bern, Sidlerstrasse 5, 3012 Bern, Switzerland}
\address[Frascati]{INFN Laboratori Nazionali di Frascati, 00044 Frascati RM, Italy}

\begin{abstract}
  Precision tests of first-row unitarity of the Cabibbo--Kobayashi--Maskawa matrix currently display two intriguing tensions, both at the $3\sigma$ level. First, combining determinations of $V_{ud}$ from superallowed $\beta$ decays with $V_{us}$ from kaon decays suggests a deficit in the unitarity relation. At the same time, a tension of similar significance has emerged between $K_{\ell 2}$ and $K_{\ell 3}$ decays. In this Letter, we point out that a measurement of the $K_{\mu3}/K_{\mu 2}$ branching fraction at the level of $0.2\%$ would have considerable impact on clarifying the experimental situation in the kaon sector, especially in view of tensions in the global fit to kaon data as well as the fact that the $K_{\mu2}$ channel is currently dominated by a single experiment. Such a measurement, as possible for example at NA62, would further provide important constraints on physics beyond the Standard Model, most notably on the role of right-handed vector currents.        
\end{abstract}

\end{frontmatter}

\thispagestyle{firstpage}

\section{Introduction}
\label{sec:intro}

Unitarity of the Cabibbo--Kobayashi--Maskawa (CKM) matrix~\cite{Cabibbo:1963yz,Kobayashi:1973fv} has a long tradition as a precision test of the Standard Model (SM). In particular, the first-row unitarity relation,
\beq
\label{unitarity}
|V_{ud}|^2+|V_{us}|^2+|V_{ub}|^2=1,
\eeq
can be probed with high precision, from a combination of $\beta$ and kaon decays that allow one to reach an uncertainty in $V_{ud}$ and $V_{us}$ of a few times $10^{-4}$. Given that $|V_{ub}|^2\simeq 1.5\times 10^{-5}$~\cite{ParticleDataGroup:2022pth}, its role can be largely ignored, and the challenge in testing Eq.~\eqref{unitarity}
lies in precision determinations of $V_{ud}$ and $V_{us}$.  

For $V_{ud}$,  superallowed nuclear $\beta$ decays ($0^+\to 0^+$ transitions) have long been the primary source of information, reaching an experimental sensitivity of $1.1\times 10^{-4}$ on $V_{ud}$~\cite{Hardy:2020qwl}. This makes nuclear corrections to the SM prediction the main source of uncertainty. In the recent literature, the discussion has focused on universal corrections from $\gamma W$ box diagrams~\cite{Marciano:2005ec,Seng:2018yzq,Seng:2018qru,Czarnecki:2019mwq,Seng:2020wjq,Hayen:2020cxh,Shiells:2020fqp} that apply equally to the nuclear case, i.e., to superallowed $\beta$ decays, as well as to neutron decay. A comparative review of these corrections is provided in~\ref{app:radiative}, leading to the values of the respective corrections in Eq.~\eqref{DeltaR} that we will use in the following. Employing the same input as Ref.~\cite{Hardy:2020qwl}  for all other corrections, this yields
\beq
\label{0+}
V_{ud}^{0^+\to 0^+}=0.97367(11)_\text{exp}(13)_{\Delta_V^R}(27)_\text{NS}[32]_\text{total},
\eeq
where the third, nuclear uncertainty from Ref.~\cite{Gorchtein:2018fxl} has also been adopted in Refs.~\cite{Hardy:2020qwl,ParticleDataGroup:2022pth}. Keeping this additional nuclear uncertainty seems warranted also in view of concerns regarding isospin-breaking corrections~\cite{Miller:2008my,Miller:2009cg,Condren:2022dji,Seng:2022epj,Crawford:2022yhi}, but improving these nuclear-structure uncertainties may be possible in the future given recent advances in ab-initio theory for nuclear $\beta$ decays~\cite{Gysbers:2019uyb,Martin:2021bud,Glick-Magid:2021xty}.

An alternative determination of $V_{ud}$ is possible from neutron decay~\cite{Czarnecki:2018okw}. This option is free of nuclear uncertainties but requires knowledge of the neutron to proton axial current matrix element. The master formula in this case thus requires information on the neutron lifetime $\tau_n$ and, in addition, on the nucleon isovector axial charge $\lambda = g_A/g_V$, which at the relevant precision is extracted from experimental measurements of the $\beta$ asymmetry in polarized neutron decay. 
With current world averages~\cite{ParticleDataGroup:2022pth}, one has
\beq
\label{n_PDG}
V_{ud}^\text{n, PDG}=0.97441(3)_f(13)_{\Delta_R}(82)_\lambda(28)_{\tau_n}[88]_\text{total},
\eeq
where the first error arises from the propagation of the uncertainty in the phase-space factor $f=1.6887(1)$~\cite{Czarnecki:2018okw}. However, especially the value of $\lambda$ carries an inflated uncertainty due to scale factors, and we believe that the current best experiments imply more information than suggested by the global averages. Therefore, using only Ref.~\cite{UCNt:2021pcg} for $\tau_n$ and Ref.~\cite{Markisch:2018ndu} for $\lambda$, we find
\beq
\label{n_best}
V_{ud}^\text{n, best}=0.97413(3)_f(13)_{\Delta_R}(35)_\lambda(20)_{\tau_n}[43]_\text{total},
\eeq
which is getting close to the sensitivity of superallowed $\beta$ decays~\eqref{0+} if there the nuclear-structure uncertainties are included. In the following, we will focus on Eqs.~\eqref{0+} and \eqref{n_best} when discussing the state of CKM unitarity, as well as their combination, 
\beq
\label{comb}
V_{ud}^\beta=0.97384(26),
\eeq
as the current most optimistic determination (to good approximation, both numbers can be considered uncorrelated, since the errors are dominated by nuclear-structure corrections and neutron-decay measurements, respectively). For completeness, we also mention the result from pion $\beta$ decay~\cite{Pocanic:2003pf,Cirigliano:2002ng,Czarnecki:2019iwz,Feng:2020zdc}
\beq
V_{ud}^\pi=0.97386(281)_\text{BR}(9)_{\tau_\pi}(14)_{\Delta_\text{RC}^{\pi\ell}}(28)_{I_{\pi\ell}}[283]_\text{total},
\eeq
with uncertainty entirely dominated by the branching fraction~\cite{Pocanic:2003pf} (the subleading errors refer to the pion lifetime $\tau_\pi$, radiative corrections~\cite{Cirigliano:2002ng,Feng:2020zdc}, and the phase-space factor, whose uncertainty mainly arises from the pion mass difference). A competitive determination requires a dedicated experimental campaign, as planned at the PIONEER experiment~\cite{PIONEER:2022yag}.

\begin{figure}[t]
	\centering
	\includegraphics[width=\linewidth,clip]{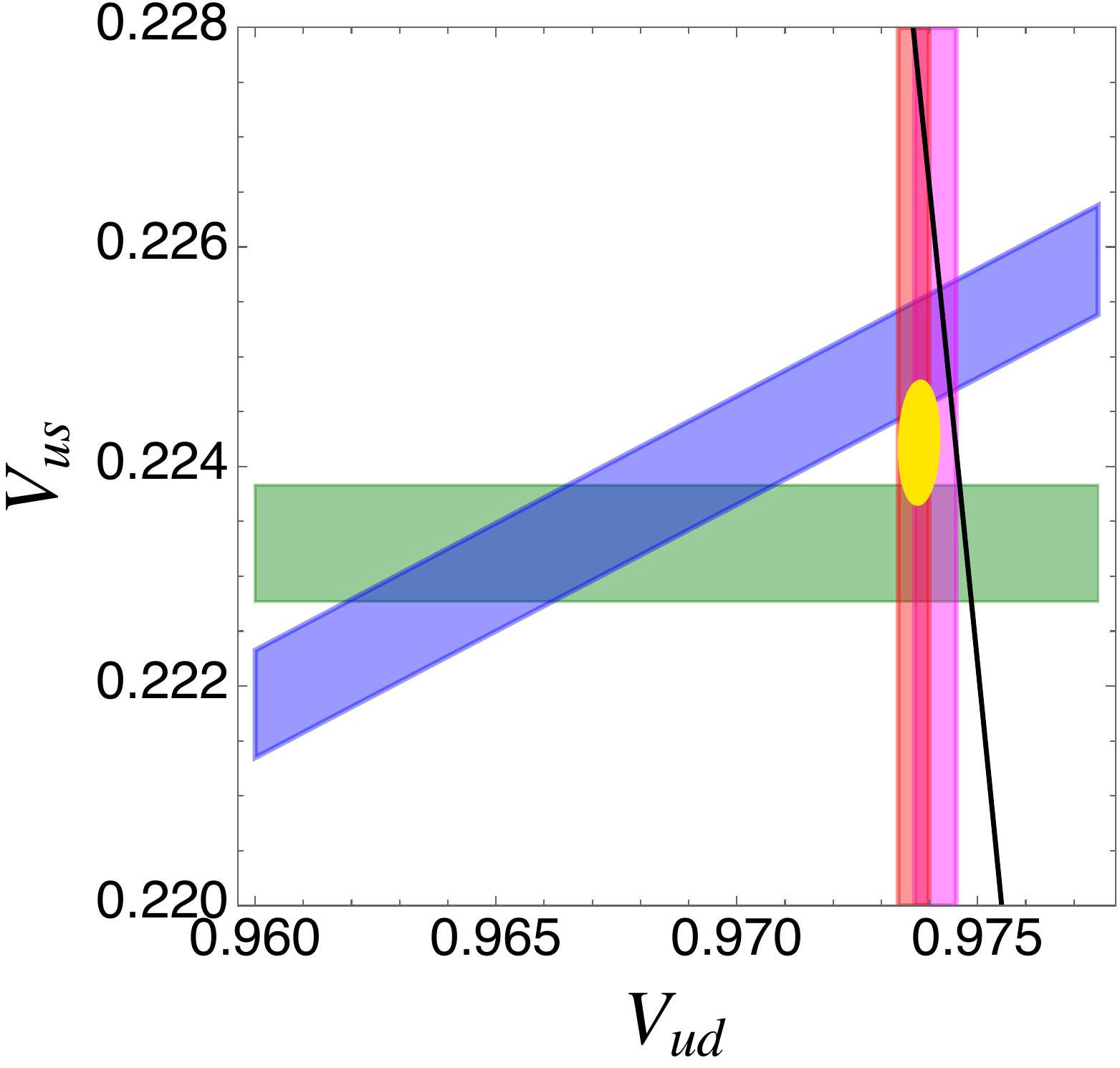}
	\caption{Constraints in the $V_{ud}$--$V_{us}$ plane. 
	The partially overlapping vertical bands correspond to $V_{ud}^{0^+\to 0^+}$ (leftmost, red) and $V_{ud}^\text{n, best}$ (rightmost, violet).
	The horizontal band (green) corresponds to $V_{us}^{K_{\ell3}}$.
	The diagonal band (blue) corresponds to 
	$({V_{us}}/{V_{ud}})_{K_{\ell 2}/\pi_{\ell 2}}$. 
	The unitarity circle is denoted by the black solid line. 
	The $68\%$ C.L.\ ellipse from a fit  to all four constraints is depicted in yellow  
	($V_{ud} = 0.97378(26)$, $V_{us}=0.22422(36)$,	$\chi^2/\text{dof} =6.4/2 $, $p$-value $4.1\%$), it deviates from the unitarity line by $2.8\sigma$. Note that the significance tends to increase in case $\tau$ decays are included.}
	\label{fig:VudVus}
\end{figure}

The best information on $V_{us}$ comes from kaon decays, $K_{\ell 2}= K\to \ell \nu_\ell$ and $K_{\ell 3}=K\to\pi \ell \nu_\ell$. The former is typically analyzed by normalizing to $\pi_{\ell 2}$ decays~\cite{Marciano:2004uf}, leading to a constraint on $V_{us}/V_{ud}$, while $K_{\ell 3}$ decays give direct access to $V_{us}$ when the corresponding form factor is provided from lattice QCD~\cite{Aoki:2021kgd}. Details of the global fit to kaon decays, as well as the input for  decay constants, form factors, and radiative corrections, are discussed in Sec.~\ref{sec:global}, leading to
\begin{align}
\label{V_us_current}
\frac{V_{us}}{V_{ud}}\bigg|_{K_{\ell 2}/\pi_{\ell 2}}&=0.23108(23)_\text{exp}(42)_{F_K/F_\pi}(16)_\text{IB}[51]_\text{total},\notag\\
V_{us}^{K_{\ell3}}&=0.22330(35)_\text{exp}(39)_{f_+}(8)_\text{IB}[53]_\text{total},
\end{align}
where the errors refer to experiment, lattice input for the matrix elements, and isospin-breaking corrections, respectively. Together with the constraints on $V_{ud}$, these bands give rise to the situation depicted in Fig.~\ref{fig:VudVus}: on the one hand, there is a tension between the best fit  and CKM unitarity, but another 
tension, arising entirely from meson decays, 
is due to the fact that the $K_{\ell 2}$ and $K_{\ell 3}$ constraints intersect away from the unitarity circle. Additional information on $V_{us}$ can be derived from $\tau$ 
decays~\cite{Gamiz:2002nu,Gamiz:2004ar}, 
but given the larger errors~\cite{HFLAV:2022pwe,Cirigliano:2021yto} we will 
continue to focus on the kaon sector.  

The main point of this Letter is that given the various tensions in the $V_{ud}$--$V_{us}$ plane, there is urgent need for additional information on the compatibility of $K_{\ell 2}$ and $K_{\ell 3}$ data, especially when it comes to interpreting either of the tensions (CKM unitarity and $K_{\ell 2}$ versus $K_{\ell 3}$) in terms of physics beyond the SM (BSM). In particular, the data base for $K_{\ell 2}$ is completely dominated by a single experiment~\cite{KLOE:2005xes},  and at the same time the global fit to all kaon data displays a relatively poor fit quality. All these points could be scrutinized by a new measurement of the $K_{\mu 3}/K_{\mu 2}$ branching fraction at the level of a few permil, as possible at the NA62 experiment. 
Further, once the experimental situation is clarified, more robust interpretations of the ensuing tensions will be possible, especially regarding the role of right-handed currents both in the strange and non-strange sector.
To make the case for the proposed measurement of the $K_{\mu 3}/K_{\mu 2}$ branching fraction, we first discuss in detail its impact on the global fit to kaon data and the implications for CKM unitarity in Sec.~\ref{sec:global}. The consequences for physics beyond the SM  are addressed
in Sec.~\ref{sec:BSM}, before we conclude in Sec.~\ref{sec:conclusions}.

\section{Global fit to kaon data and implications for  CKM unitarity}
\label{sec:global}

\begin{table*}[t]
	\centering
	\renewcommand{\arraystretch}{1.3}
	\scalebox{0.87}{
	\begin{tabular}{l r r r r r rr}
	\toprule
	& current fit & \multicolumn{3}{c}{$K_{\mu 3}/K_{\mu 2}$ BR at $0.5\%$} & \multicolumn{3}{c}{ $K_{\mu 3}/K_{\mu 2}$ BR  at $0.2\%$}\\
	& & central & $+2\sigma$ & $-2\sigma$ & central & $+2\sigma$ & $-2\sigma$\\\midrule
	$\chi^2/\text{dof}$ & $25.5/11$ & $25.5/12$ & $31.8/12$ & $32.1/12$ & $25.5/12$ & $35.6/12$ & $35.9/12$ \\
	$p$-value [\%] & $0.78$ & $1.28$ & $0.15$ & $0.13$ & $1.28$ & $0.04$ & $0.03$\\\midrule
	BR($\mu\nu$) [\%]& $63.58(11)$ & $63.58(09)$ & $63.44(10)$ & $63.72(11)$ & $63.58(08)$ & $63.36(10)$ & $63.80(11)$\\
	$S(\mu\nu)$ & $1.1$ & $1.1$ & $1.3$ & $1.4$ & $1.2$ & $1.6$ & $1.7$\\
	BR($\pi\pi^0$) [\%] & $20.64(7)$ & $20.64(6)$& $20.73(7)$& $20.55(8)$& $20.64(6)$& $20.78(7)$& $20.50(10)$\\
	$S(\pi\pi^0)$ & $1.1$ & $1.2$ & $1.3$ & $1.5$ & $1.2$ & $1.5$ & $2.0$\\ 
	BR($\pi\pi\pi$) [\%] & \multicolumn{7}{c}{$5.56(4)$} \\ 
	$S(\pi\pi\pi)$ & \multicolumn{7}{c}{$1.0$}\\
	BR($K_{e3}$) [\%] & $5.088(27)$ & $5.088(24)$ & $5.113(25)$ & $5.061(31)$ & $5.088(23)$ & $5.128(24)$ & $5.046(32)$\\
	$S(K_{e3})$ & $1.2$ & $1.2$ & $1.2$ & $1.6$ & $1.3$ & $1.3$ & $1.8$\\
	BR($K_{\mu3}$) [\%] & $3.366(30)$ & $3.366(13)$ & $3.394(16)$ & $3.336(27)$ & $3.366(7)$ & $3.411(13)$ & $3.320(18)$\\
	$S(K_{\mu3})$ & $1.9$ & $1.2$ & $1.5$ & $2.6$ & $1.1$ & $2.2$ & $3.1$\\
	BR($\pi\pi^0\pi^0$) [\%] & \multicolumn{7}{c}{$1.764(25)$}\\ 
	$S(\pi\pi^0\pi^0)$ & \multicolumn{7}{c}{$1.0$}\\
	$\tau_\pm$ [ns] & $12.384(15)$ & $12.384(15)$ & $12.382(15)$ & $12.385(15)$ & $12.384(15)$ & $12.381(15)$ & $12.386(15)$\\
	$S(\tau_\pm)$ & \multicolumn{7}{c}{$1.2$}\\\midrule
	$\frac{V_{us}}{V_{ud}}\Big|_{K_{\ell 2}/\pi_{\ell 2}}$ & 
	$0.23108(51)$ & $0.23108(50)$ & $0.23085(51)$ & $0.23133(51)$ & $0.23108(49)$ & $0.23071(51)$ & $0.23147(52)$\\
	$V_{us}^{K_{\ell3}}$ & $0.22330(53)$ & $0.22337(51)$ & $0.22360(52)$ & $0.22309(54)$ & $0.22342(49)$ & $0.22386(52)$ & $0.22287(52)$\\
	$\frac{F_K}{F_\pi}\frac{V_{us}}{V_{ud}}\Big|_{K_{\ell 2}/\pi_{\ell 2}}$ & 
	$0.27679(34)$ & $0.27679(31)$ & $0.27651(35)$ & $0.27709(34)$ & $0.27679(30)$ & $0.27634(33)$ & $0.27726(35)$\\
	$f_+(0)V_{us}^{K_{\ell3}}$ & $0.21656(35)$ & $0.21662(31)$ & $0.21685(33)$ & $0.21636(35)$ & $0.21667(28)$ & $0.21710(32)$ & $0.21614(34)$\\\midrule
	\multirow{2}{*}{$\Delta_\text{CKM}^{(1)}$} & $-0.00176(56)$ & $-0.00173(55)$ & $-0.00162(56)$ & $-0.00185(56)$ & $-0.00171(55)$ & $-0.00151(56)$ & $-0.00195(56)$\\
	& 
	$-3.1\sigma$ & $-3.1\sigma$ & $-2.9\sigma$ & $-3.3\sigma$ & $-3.1\sigma$ & $-2.7\sigma$ & $-3.5\sigma$\\
	\multirow{2}{*}{$\Delta_\text{CKM}^{(2)}$} & $-0.00098(58)$ & $-0.00098(58)$ & $-0.00108(58)$ & $-0.00087(58)$ & $-0.00098(58)$ & $-0.00114(58)$ & $-0.00081(58)$\\
	&
	$-1.7\sigma$ & $-1.7\sigma$ & $-1.9\sigma$ & $-1.5\sigma$ & $-1.7\sigma$ & $-2.0\sigma$ & $-1.4\sigma$\\
	\multirow{2}{*}{$\Delta_\text{CKM}^{(3)}$} & $-0.0164(63)$ & $-0.0157(60)$ & $-0.0118(62)$ & $-0.0202(63)$ & $-0.0153(59)$ & $-0.0083(62)$ & $-0.0233(62)$\\
	& 
	$-2.6\sigma$ & $-2.6\sigma$ & $-1.9\sigma$ & $-3.2\sigma$ & $-2.6\sigma$ & $-1.4\sigma$ & $-3.8\sigma$\\
\bottomrule
	\end{tabular}
	}
	\caption{Fit results for the current global fit as well as variants including a new measurement of the $K_{\mu3}/K_{\mu 2}$ branching fraction, with uncertainty of $0.5\%$ and $0.2\%$, respectively, and central value either as expected from the current fit, BR($K_{\mu3}$)/BR($K_{\mu2})=0.05294(51)$, or shifted by $\pm2\sigma$ of the current fit error. In each channel, the scale factors are given to quantify the tension as originating therefrom~\cite{ParticleDataGroup:2022pth}. Note that the branching ratios for $\pi\pi\pi$ and $\pi\pi^0\pi^0$ are virtually unaffected by the new measurement due to very few correlated ratios with the (semi-) leptonic channels in the data base (in cases in which no significant changes occur, only a single entry is given that applies to all columns). 
	The values of $V_{us}$ and $V_{us}/V_{ud}$ are extracted using the same input as described in the main text, adding in quadrature all uncertainties given in Eq.~\eqref{V_us_current}. $\Delta_\text{CKM}^{(1,2,3)}$ are defined in Eq.~\eqref{Delta_CKM}, and $\Delta_\text{CKM}^{(1,2)}$ are evaluated using $V_{ud}^\beta$ from Eq.~\eqref{comb}.}
	\label{tab:kaon_fit}
\end{table*}

The current values for $V_{us}$ and $V_{us}/V_{ud}$ given in Eq.~\eqref{V_us_current} are obtained from a global fit to kaon decays~\cite{FlaviaNetWorkingGrouponKaonDecays:2010lot,Moulson:2013wi,Moulson:2014cra,Moulson:2017ive}, updated to include the latest measurements, radiative corrections, and hadronic matrix elements.  In particular, the fit includes data on $K_S$ decays from Refs.~\cite{KLOE:2006jcl,KLOE:2006vvm,Batley:2007zzb,KTeV:2010sng,KLOE:2010yit,KLOE-2:2019rev,KLOE-2:2019fdf}, on $K_L$ decays from Refs.~\cite{Vosburgh:1971zk,E731:1992nnl,KTeV:2000avq,KLOE:2003pmf,Lai:2002sr,KTeV:2004hpx,NA48:2004utd,KLOE:2005lau,KLOE:2005vdt,KLOE:2006ane,KTeV:2006diq,NA48:2006jeq}, and on charged-kaon decays from Refs.~\cite{Fitch:1965zz,Auerbach:1967nip,Ott:1971rs,Vaisenberg:1976tz,Usher:1992pz,Koptev:1995je,KEK-E246:2001mzq,Sher:2003fb,KLOE:2003new,KLOE:2005xes,NA482:2006vnw,KLOE:2007wlh,KLOE:2007jte,KLOE:2008tnn,KLOEKLOE-2:2014tsu}. Since we focus on the impact of a new $K_{\mu 3}/K_{\mu 2}$ measurement, e.g., at NA62, we reproduce the details of the charged kaon fit in Table~\ref{tab:kaon_fit}, where, however, the value for $V_{us}$ from $K_{\ell3}$ decays includes the results obtained for all decay modes, accounting for correlations among them.  
The extraction of $V_{us}$ from $K_{\ell 3}$ decays requires further input on the respective form factors, which are taken in the dispersive parameterization from Ref.~\cite{Bernard:2009zm}, constrained by data from Refs.~\cite{Yushchenko:2003xz,Yushchenko:2004zs,NA48:2004jcz,KTeV:2004ozu,KLOE:2006kms,KLOE:2007vlu,NA482:2018rgv}. This leaves form-factor normalizations, decay constants, and isospin-breaking corrections in both $K_{\ell 2}$ and $K_{\ell 3}$ decays. 

For $K_{\ell 2}$ we follow the established convention to consider the ratio to $\pi_{\ell 2}$ decays~\cite{Marciano:2004uf} (pion lifetime~\cite{Eckhause:1965zz,Nordberg:1967zz,Ayers:1971kz,Dunaitsev:1972st,Koptev:1995je,Numao:1995qf} and branching fraction~\cite{Bryman:1985bv,Britton:1992pg,Czapek:1993kc,PiENu:2015seu} are taken from Ref.~\cite{ParticleDataGroup:2022pth}), since in this ratio certain structure-dependent radiative corrections~\cite{Knecht:1999ag,Cirigliano:2011tm} cancel and only the ratio of decay constants $F_K/F_\pi$ needs to be provided. We use the isospin-breaking corrections from Ref.~\cite{DiCarlo:2019thl} together with the $N_f=2+1+1$ isospin-limit ratio of decay constants $F_K/F_\pi=1.1978(22)$~\cite{Dowdall:2013rya,Bazavov:2017lyh,Miller:2020xhy,ExtendedTwistedMass:2021qui}, where this average accounts for statistical and systematic correlations between the results, some of which make use of the same lattice ensembles. For $K_{\ell 3}$ decays we use the radiative corrections from Refs.~\cite{Seng:2021boy,Seng:2021wcf,Seng:2022wcw} (in line with the earlier calculations~\cite{Cirigliano:2001mk,Cirigliano:2008wn}), the strong isospin-breaking correction $\Delta_\text{SU(2)} = 0.0252(11)$ from Refs.~\cite{Gasser:1984ux,Cirigliano:2001mk} evaluated with the $N_f=2+1+1$ quark-mass double ratio $Q=22.5(5)$ and ratio $m_s/m_{ud}=27.23(10)$, both from Ref.~\cite{Aoki:2021kgd} (the value of $Q$ is consistent with $Q=22.1(7)$ from $\eta\to 3\pi$~\cite{Colangelo:2018jxw} and $Q=22.4(3)$ from the Cottingham approach~\cite{Stamen:2022uqh}), and the form-factor normalization $f_+(0)=0.9698(17) $~\cite{Aoki:2021kgd,Carrasco:2016kpy,FermilabLattice:2018zqv}.\footnote{We use the $N_f=2+1+1$ average from Ref.~\cite{Aoki:2021kgd}, which is dominated by Ref.~\cite{FermilabLattice:2018zqv}. This value is in agreement with the $N_f=2+1$ average $f_+(0)=0.9677(27)$~\cite{Aoki:2021kgd,Bazavov:2012cd,RBCUKQCD:2015joy}, and also marginally with Ref.~\cite{Ishikawa:2022ulx} due to the large asymmetric error.} This global fit then defines the current baseline result given in Eq.~\eqref{V_us_current} and the first column of Table~\ref{tab:kaon_fit}. The resulting constraints in the $V_{ud}$--$V_{us}$ plane are shown in Fig.~\ref{fig:VudVus}, illustrating the tensions among the two kaon bands and the $V_{ud}$ determinations from superallowed $\beta$ and neutron decays (commonly denoted by  $V_{ud}^\beta$, see Eq.~\eqref{comb}). To quantify the tensions with CKM unitarity, there are thus several possible variants
\begin{align}
\label{Delta_CKM}
 \Delta_\text{CKM}^{(1)}&=\big|V_{ud}^\beta\big|^2+\big|V_{us}^{K_{\ell 3}}\big|^2-1,\notag\\
 \Delta_\text{CKM}^{(2)}&=\big|V_{ud}^\beta\big|^2+\big|V_{us}^{K_{\ell 2}/\pi_{\ell 2},\, \beta}\big|^2-1,\notag\\
 \Delta_\text{CKM}^{(3)}&=\big|V_{ud}^{K_{\ell 2}/\pi_{\ell 2},\, K_{\ell 3}}\big|^2+\big|V_{us}^{K_{\ell 3}}\big|^2-1,
\end{align}
which differ in their BSM interpretation; see Sec.~\ref{sec:BSM}. The numerical results are also included in Table~\ref{tab:kaon_fit}.  

With the results shown in Table~\ref{tab:kaon_fit}, the impact of a new $K_{\mu 3}/K_{\mu 2}$ measurement becomes apparent: first, the biggest scale factor in the current fit arises in the $K_{\mu 3}$ channel, while the $K_{\mu 2}$ data base is dominated by a single experiment~\cite{KLOE:2005xes}. To date, no direct measurements of the ratio exist apart from the very early ones in Refs.~\cite{Auerbach:1967nip,Garland:1968zz,Zeller:1969cy}, which, however, are not precise enough to be included in the fit, in such a way that the current value quoted in Table~\ref{tab:kaon_fit} derives in a complicated way from the correlations in the global fit. 

Accordingly, new information on these two channels is paramount to resolve or corroborate current tensions, which lead to a $p$-value below $1\%$. The expected impact is then illustrated by the benchmarks in the subsequent columns, showing the result of the global fit when adding a hypothetical measurement of the $K_{\mu 3}/K_{\mu 2}$ branching fraction with a precision of $0.5\%$ and $0.2\%$, and central values as expected from the current fit or shifted by $2\sigma$ of its error in either direction. The 0.5\% and 0.2\% precision benchmarks were chosen because they are within experimental reach. The main goal of the NA62 experiment at the CERN SPS is to measure the branching ratio for the rare decay $K^+\to\pi^+\nu\bar{\nu}$; to this end, the experiment normally runs at high intensity with a dedicated trigger. In order to perform high-precision measurements of branching-fraction ratios relevant to first-row CKM unitarity tests, the experiment could collect the needed data in a short run with a minimum-bias trigger, at low intensity and with special emphasis on maintaining data-taking conditions as stable as possible. It is likely that with a run of less than two weeks in duration, the uncertainty on the measurement of $K_{\mu 3}/K_{\mu 2}$ would be dominated by systematics. Estimation of NA62's systematic sensitivity in this channel is beyond the scope of this work; we limit ourselves to the observation that the experiment has high-precision tracking for both beam and secondary particles, redundant particle-identification systems, and excellent calorimetry and photon detection. Although NA62 to date has focused on searches for rare decays rather than precision measurements of abundant channels, in 2013, with the NA48/2 setup even before experimental upgrades to the systems described above, the collaboration measured the ratio of branching fractions $K_{e2}/K_{\mu2}$ to within 0.4\%~\cite{NA62:2012lny}. More recently, NA62 measured the branching fraction for the rare decay $K^+\to\pi^+\mu^+\mu^-$ to be $9.15(8)\times10^{-8}$, with a systematic contribution to the uncertainty of just 0.32\%~\cite{NA62:2022qes}. We therefore consider 0.5\% and 0.2\% as relatively straightforward and cautiously optimistic benchmarks for an experiment like NA62, noting in particular that while NA62 is to our knowledge the only running experiment able to make incisive measurements relevant to first-row CKM unitarity tests, such measurements are clearly an experimental possibility and could be proposed at other facilities. 

Turning our attention back to Table~\ref{tab:kaon_fit}, we see that in case of a $2\sigma$ shift, already a $0.5\%$ measurement would significantly affect the scale factors and alter the extracted CKM matrix elements by almost $0.5\sigma$, an effect that would increase further for the $0.2\%$ scenario. In this case, the significance of the tension in $\Delta_\text{CKM}^{(3)}$, the measure directly derived from kaon decays, would increase or decrease by more than $1\sigma$, demonstrating that a new precision measurement of the $K_{\mu 3}/K_{\mu 2}$ branching fraction really has the potential to either resolve or substantially corroborate the tension between the $K_{\ell 2}$ and $K_{\ell 3}$ CKM-element determinations. Once the experimental situation in the kaon sector is clarified, possible BSM interpretations become much more robust, as we discuss in the subsequent section.

\section{Constraints on physics beyond the Standard Model}
\label{sec:BSM}

The current tension with CKM unitarity has triggered renewed interest in possible BSM explanations~\cite{Belfatto:2019swo,Coutinho:2019aiy}, including interpretations in terms of vector-like quarks~\cite{Cheung:2020vqm,Belfatto:2021jhf,Branco:2021vhs} and leptons~\cite{Crivellin:2020ebi,Kirk:2020wdk}, as modifications of the Fermi constant~\cite{Marciano:1999ih,Crivellin:2021njn},  in the context of lepton flavor universality~\cite{Crivellin:2020lzu,Crivellin:2020klg,Capdevila:2020rrl,Crivellin:2021sff,Crivellin:2020oup,Marzocca:2021azj}, and even allowing for a correlation with di-electron searches at the LHC~\cite{Crivellin:2021rbf,Crivellin:2021bkd}.
Here, we illustrate the impact of our proposed $K_{\mu 3}/K_{\mu 2}$ measurement via the constraints on right-handed currents~\cite{Bernard:2007cf,Alioli:2017ces,Grossman:2019bzp,Cirigliano:2021yto}, which can not only address the tension between $\beta$ and kaon decays, but also between $K_{\ell 2}$ and $K_{\ell 3}$. This discussion becomes most transparent in terms of the $\Delta_\text{CKM}^{(i)}$ introduced in Eq.~\eqref{Delta_CKM}.
 
\begin{figure}[t]
	\centering
	\includegraphics[width=\linewidth,clip]{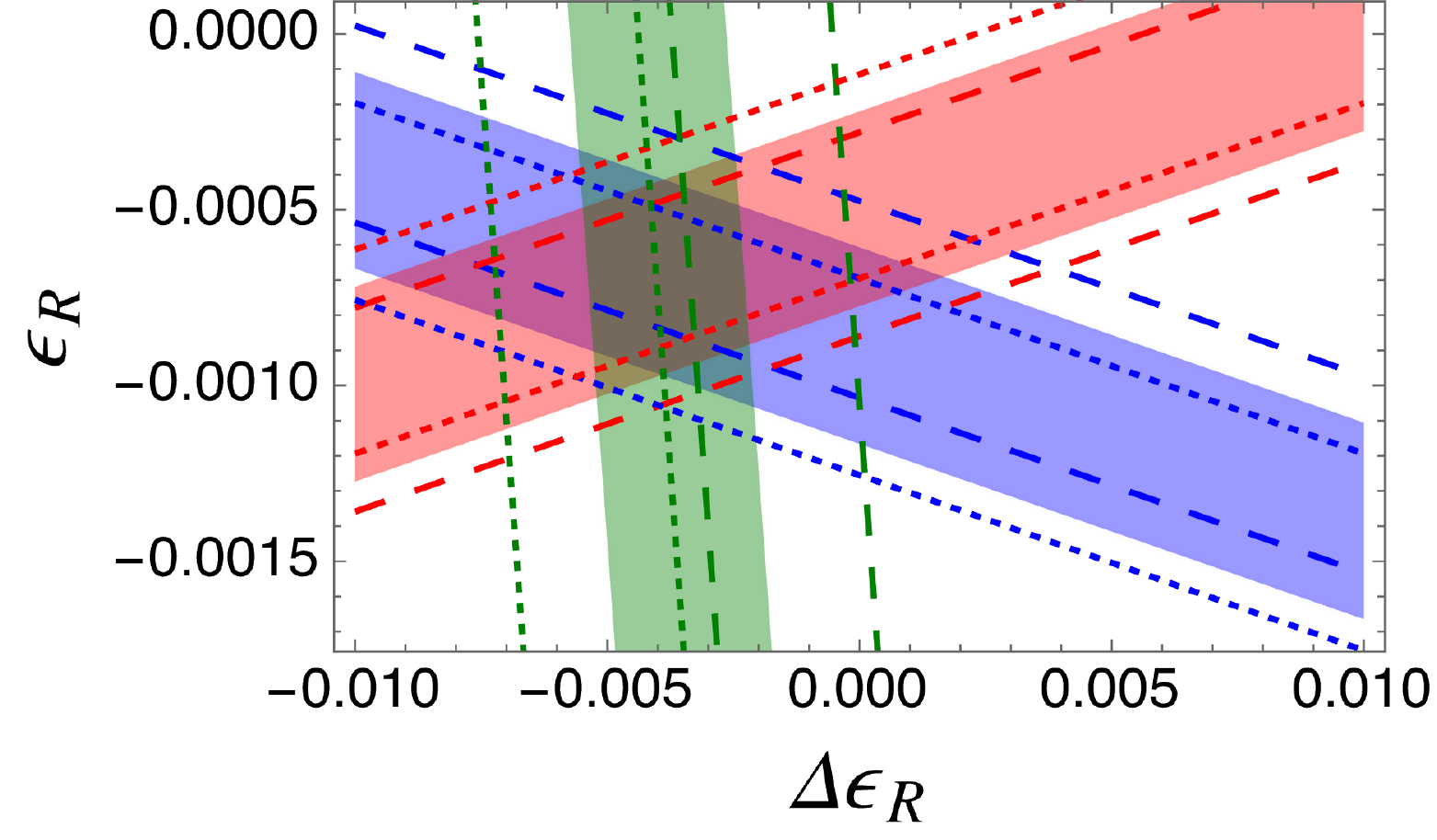}
	\caption{Constraints in the $\Delta \eps_R$--$\eps_R$ plane from the $\Delta_\text{CKM}^{(i)}$ introduced in Eq.~\eqref{Delta_CKM}. 
	The bands with positive slope (red) correspond to $\Delta_\text{CKM}^{(2)}$. The bands with small negative slope (blue) correspond to $\Delta_\text{CKM}^{(1)}$, 
	while the bands with steep negative slope (green) correspond to $\Delta_\text{CKM}^{(3)}$. The filled bands reflect the current situation~\eqref{eps}, the long-dashed ones the $+2\sigma$ scenario~\eqref{epsp2}, and the short-dashed ones the opposite case~\eqref{epsm2}. 
	Note that in each case the three bands essentially overlap by construction, since $V_{ud}$, $V_{us}$, subject to the unitarity constraint, and the BSM contamination via $\Delta \eps_R$, $\eps_R$, amount to three free parameters. The main impact of the proposed new measurement of the $K_{\mu 3}/K_{\mu 2}$ branching fraction thus concerns a corresponding shift in the $\Delta_\text{CKM}^{(3)}$ band if the $\pm 2\sigma$ scenarios  were realized.} 
	\label{fig:epsR}
\end{figure}

In general, a single parameter is not sufficient to explain both tensions, as they are governed by a-priori independent operators, and we therefore introduce two parameters $\eps_R$, $\eps_R^{(s)}$ (or equivalently $\eps_R$ and $\Delta \eps_R \equiv \eps_R^{(s)} - \eps_R$, normalized as in Ref.~\cite{Cirigliano:2021yto}) to quantify right-handed currents in the non-strange and strange sectors, respectively.  Working at first order in $\eps$, the CKM elements in Eq.~\eqref{Delta_CKM} as extracted from the (vector-current mediated) three-particle decays are contaminated by $1+\eps$, the ones from the (axial-current mediated) two-particle decays by $1-\eps$, resulting in 
  \begin{align}
  \Delta_\text{CKM}^{(1)}&=2\eps_R+2\Delta \eps_R V_{us}^2,\notag\\
  \Delta_\text{CKM}^{(2)}&=2\eps_R-2\Delta \eps_R V_{us}^2,\notag\\
  \Delta_\text{CKM}^{(3)}&=2\eps_R+2\Delta \eps_R\big(2-V_{us}^2\big). 
 \end{align}
 The corresponding constraints are shown in Fig.~\ref{fig:epsR} 
 and point  to  non-zero values for both $\eps_R$ and $\Delta \eps_R$.
  $\eps_R$ can be isolated by taking the average of 
 $\Delta_\text{CKM}^{(1)}$ and $\Delta_\text{CKM}^{(2)}$, 
 while $\Delta \eps_R$ is obtained from the combination 
 \begin{equation}
 r  \equiv \left( \frac{1 +  \Delta_\text{CKM}^{(2)}}{1 + \Delta_\text{CKM}^{(3)} } \right)^{1/2} = 
 \frac{\frac{V_{us}}{V_{ud}}\Big|_{K_{\ell 2}/\pi_{\ell 2}}}{\frac{V_{us}^{K_{\ell 3}}}{V_{ud}^\beta}}=1-2\Delta\eps_R.
 \end{equation}
 Using current input from  Eqs.~\eqref{comb} and~\eqref{V_us_current}, one obtains: 
 \begin{align}
 \label{eps}
  \eps_R &=-0.69(27)\times 10^{-3} && [2.5 \sigma],
  \notag \\
 \Delta\eps_R &=-3.9(1.6)\times 10^{-3} && [2.4\sigma].
 \end{align}
With a projected measurement of the $K_{\mu 3}/K_{\mu 2}$ branching ratio at 0.2\% level 
at $2\sigma$ above the current measurement, the above numbers change to
  \begin{align}
  \label{epsp2}
  \eps_R &=-0.67(27)\times 10^{-3} &&[2.5 \sigma],
  \notag\\
 \Delta\eps_R &=-1.8(1.6)\times 10^{-3} && [1.1\sigma],
 \end{align}
while a future  measurement at 0.2\% with central value $2\sigma$ below 
the current one would give
 \begin{align}
  \label{epsm2}
  \eps_R &=-0.70(27)\times 10^{-3} && [2.6 \sigma],
  \notag \\
 \Delta\eps_R &=-5.7(1.6)\times 10^{-3} && [3.5\sigma].
 \end{align}
This shows that the proposed measurement would have a significant impact on revealing or further constraining right-handed charged currents involving strange quarks.  
In particular, the non-vanishing value of $\eps_R$ is mainly driven by the $\beta$-decay observables, while the goal of the new $K_{\mu 3}/K_{\mu 2}$ input would be a conclusive answer  to the question whether or not further strangeness right-handed currents need to be invoked. Here, the sensitivity of $\Delta\eps_R$ to the different scenarios  reflects similar changes in $\Delta_\text{CKM}^{(3)}$ as observed in Table~\ref{tab:kaon_fit}.

We note here that other probes of $\eps_R$ and $\Delta \eps_R$ are currently less constraining and are not reported in  Fig.~\ref{fig:epsR}.
In particular, $\epsilon_R$ can be determined from the comparison of 
the experimentally measured axial charge $\lambda = g_A/g_V$ and its 
value computed in lattice QCD~\cite{Aoki:2021kgd,Chang:2018uxx,Gupta:2018qil}, up to a recently uncovered electromagnetic correction~\cite{Cirigliano:2022hob}. This results in $\eps_R = - 0.2(1.2)\% $.
Similarly, assuming a high-scale origin for the right-handed couplings 
and writing the operator in an SU(2) $\times$ U(1) invariant form, 
one obtains constraints from associated Higgs production at the few-percent level~\cite{Alioli:2017ces}. 

A similar analysis could be performed in terms of pseudoscalar couplings $\eps_P$, $\eps_P^{(s)}$, which only affect the axial-current induced two-particle decays via $1-\eps$. Accordingly, we have
\begin{align}
r&=1-\Delta \eps_P,\qquad \Delta\eps_P=\eps_P^{(s)}-\eps_P,\\
\Delta_\text{CKM}^{(1)}&=0,\quad \Delta_\text{CKM}^{(2)}=-2\Delta \eps_P V_{us}^2,\quad \Delta_\text{CKM}^{(3)}=2\Delta\eps_P\big(1-V_{us}^2\big),\notag
\end{align}
which shows that not all tensions can be explained in terms of a pseudoscalar operator. 

\section{Conclusions}
\label{sec:conclusions}

In this Letter, we have made the case that a new precision measurement of the $K_{\mu 3}/K_{\mu 2}$ branching fraction would have a major impact on clarifying the current tensions in the unitarity relation for the first row of the CKM matrix. As a first step, we have presented an update of the global fit to kaon data; see Eq.~\eqref{V_us_current}, including the most recent measurements, radiative corrections, and hadronic matrix elements. We have also performed a comparative review of recent calculations of the $\gamma W$ box corrections to the $\beta$-decay observables, leading to the corrections in Eq.~\eqref{DeltaR}, to obtain up-to-date values for $V_{ud}$ determined from superallowed $\beta$ decays and neutron decay; see Eq.~\eqref{comb} for the combined result when retaining only the current best neutron-decay measurements. Our main results are given in Table~\ref{tab:kaon_fit}, where we show the impact of the proposed $K_{\mu 3}/K_{\mu 2}$ measurement at $0.5\%$ and $0.2\%$ precision. These results demonstrate that already at $0.5\%$ the additional input would be valuable to decide whether the current tensions, especially between $K_{\ell 2}$ and $K_{\ell 3}$ decays, are of experimental origin or point towards BSM effects, while a $0.2\%$ measurement would have considerable impact on the global picture: the measure $\Delta_\text{CKM}^{(3)}$, which quantifies the CKM tension if only kaon-decay input is used, changes within $[-3.8\sigma,-1.4\sigma]$ when the $K_{\mu 3}/K_{\mu 2}$ branching fraction is varied by $\pm 2\sigma$ around its current expectation, and should thus provide a conclusive answer to the $K_{\ell 2}$--$K_{\ell 3}$ tension. We further studied the BSM implications together with the $\beta$-decay observables, the main conclusion being that the proposed measurement would yield a powerful constraint on contributions from right-handed currents in the strange sector. Experimentally, such a measurement would be possible at NA62, presenting a unique opportunity to resolve or corroborate current tensions in the precision test of CKM unitarity.

\section*{Acknowledgments}

We thank Joel C.\ Swallow for helpful discussions on the statistical estimates for NA62. 
Financial support from  the SNSF (Project Nos.\ PP00P21\_76884 and  PCEFP2\_181117) and  the U.S.\ DOE under Grant No.\ DE-FG02-00ER41132 is gratefully acknowledged. MH thanks the  INT at the University of Washington for its
hospitality and the DOE for partial support during a visit when part of this work was performed. 

\appendix

\section{Radiative corrections to $\boldsymbol{\beta}$ decays}
\label{app:radiative}

The determination of $V_{ud}$ from superallowed $\beta$ decays and neutron decay critically depends on radiative corrections. In particular, the universal corrections, $\Delta_R^V$ and $\Delta_R$, respectively, have been evaluated by several groups in the recent literature~\cite{Seng:2018yzq,Seng:2018qru,Czarnecki:2019mwq,Seng:2020wjq,Hayen:2020cxh,Shiells:2020fqp}, leading to significant changes with respect to the earlier evaluation from Ref.~\cite{Marciano:2005ec}. Many aspects of the different evaluations are highly correlated, in such a way that a simple average as proposed in Ref.~\cite{Hardy:2020qwl} is  not adequate. In this appendix, we review  and compare the available calculations and present our best estimate for $\Delta_R^V$ and $\Delta_R$.

The starting point is the master formula~\cite{Czarnecki:2004cw}
\begin{align}
\label{master}
 1+\Delta_R&=\bigg[1+\frac{\alpha}{2\pi}\bigg(\bar g(E_m)-3\log\frac{m_p}{2E_m}\bigg)\bigg]\\
 &\times\bigg[L(2E_m,m_p)+\frac{\alpha}{2\pi}\delta+2\Box_A^{\gamma W}\big|_{Q^2\leq Q_0^2}\bigg]\notag\\
 &\times\bigg[\tilde S(m_p,M_Z)-\frac{\alpha}{2\pi}\log\frac{M_W}{M_Z}+2\Box_A^{\gamma W}\big|_{Q^2\geq Q_0^2}+\text{NLL}\bigg]\notag
\end{align}
for the radiative correction in neutron decay, 
which involves the following contributions: $E_m=1.292581\MeV$ is the end-point electron energy, $m_p$ the proton mass, $\frac{\alpha}{2\pi}\bar g(E_m)=0.015035$ denotes the Sirlin function~\cite{Sirlin:1967zza}, $\frac{\alpha}{2\pi}\delta=-0.00043$ emerges as a Coulomb correction necessary in the factorized form~\eqref{master}, and 
\beq
L(2E_m,m_p)=1+\frac{3\alpha}{2\pi}\log\frac{m_p}{2E_m}+\Order\big(\alpha^2\big)
\eeq
resums the large logarithms below $m_p$.  Dropping all these corrections specific to neutron decay, Eq.~\eqref{master} reduces to $\Delta_R^V$, 
\begin{align}
 1+\Delta_R^V&=\bigg[1+2\Box_A^{\gamma W}\big|_{Q^2\leq Q_0^2}\bigg]\\
 &\times\bigg[\tilde S(m_p,M_Z)-\frac{\alpha}{2\pi}\log\frac{M_W}{M_Z}+2\Box_A^{\gamma W}\big|_{Q^2\geq Q_0^2}+\text{NLL}\bigg].\notag
\end{align}
Next,
\beq
\tilde S(m_p,M_Z)=1+\frac{3\alpha}{2\pi}\log\frac{M_Z}{m_p}+\Order\big(\alpha^2\big)
\eeq
resums the large logarithms above $m_p$, except for the logarithm included in 
 the $\gamma W$ box correction  $\Box_A^{\gamma W}$, because this logarithm is taken into account later by including the running of $\alpha$ in the evaluation of the DIS region. Therefore, one would incur double counting when using the original $S(m_p,M_Z)$ from Ref.~\cite{Czarnecki:2004cw}, and the corresponding change in the anomalous dimension is indicated by the tilde.  $\Box_A^{\gamma W}$
 has been the subject of most of the recent discussion in the literature. It can be expressed in the form~\cite{Seng:2018qru,Shiells:2020fqp}
\beq
\label{box_definition}
\Box_A^{\gamma W}=\frac{\alpha}{2\pi}\int_0^\infty\frac{dQ^2}{Q^2}\frac{M_W^2}{M_W^2+Q^2}\int_0^1 dx\,F_3^{(0)}\big(x,Q^2\big)\frac{1+2r}{(1+r)^2},
\eeq
with $r=\sqrt{1+4m_p^2x^2/Q^2}$ and the isoscalar structure function $F_3^{(0)}$ in the normalization of Refs.~\cite{Shiells:2020fqp,ParticleDataGroup:2022pth} (a factor $4$ larger than in Refs.~\cite{Seng:2018yzq,Seng:2018qru}). In the master formula~\eqref{master} the box contribution is separated into low- and high-$Q^2$ regions by introducing a cut $Q_0^2$ in Eq.~\eqref{box_definition}, to be able to pair each part with the appropriate resummation of logarithms. Finally, Eq.~\eqref{master} includes an estimate of next-to-leading logarithms, $\text{NLL}=-0.00010$~\cite{Czarnecki:2004cw}.

Before turning to $\Box_A^{\gamma W}$, we briefly comment on the resummation of the logarithmic corrections. For $\tilde S(m_p,M_Z)$ we take over the formalism from Ref.~\cite{Czarnecki:2004cw}, but use as starting point the $\overline{\text{MS}}$ value at the $Z$-boson mass $\hat\alpha(M_Z)^{-1}=127.952(9)$~\cite{ParticleDataGroup:2022pth}, which produces 
\beq
\tilde S(m_p,M_Z)=1.01682.
\eeq
For the resummation of the infrared logarithms, however, we do not rely on current quark masses for $q=u,d,s$~\cite{Czarnecki:2004cw}, but instead switch to a hadronic scheme below $m_p$. In fact, this strategy better matches the calculation of the anomalous dimension of the leading logarithm, which, above $m_p$, is derived from quark-level diagrams, but below relies on loop diagrams with hadronic degrees of freedom. We obtain
\begin{align}
\label{L_hadronic}
L(2E_m,m_p)&=\bigg(\frac{\hat\alpha(m_\mu)}{\hat\alpha(2E_m)}\bigg)^{9/4}
\bigg(\frac{\hat\alpha(\mpi)}{\hat\alpha(m_\mu)}\bigg)^{9/8}\bigg(\frac{\hat\alpha(\mk)}{\hat\alpha(\mpi)}\bigg)^{1}\bigg(\frac{\hat\alpha(m_p)}{\hat\alpha(\mk)}\bigg)^{9/10}\notag\\
&=1.02090,
\end{align}
where we determined $\hat \alpha$ by conversion from the on-shell scheme at $m_e$~\cite{Baikov:2012rr}. Numerically, the result is close to  $L(2E_m,m_p)=1.02094$ from Ref.~\cite{Czarnecki:2004cw} (both larger than  the leading logarithm $L(2E_m,m_p)\to 1.02054$), but the main difference is conceptual: while the scheme of Ref.~\cite{Czarnecki:2004cw} allows one to smoothly match $\hat\alpha$ from the low- and high-energy end at $m_p$, this comes at the expense of arbitrarily chosen quark masses. In our scheme, the running includes the relevant degrees of freedom in the low- and high-energy part of the calculation, with a matching correction at $m_p$ that converts the two schemes. 
Including, in addition to charged pions and kaons, also $\rho$ and $K^*$ in the resummation~\eqref{L_hadronic} only leads to marginal changes. 

\begin{table}[t]
	\centering
	\scalebox{0.759}{
	\begin{tabular}{l r r r r r r}
	\toprule
	 & \cite{Shiells:2020fqp} & \cite{Seng:2018yzq,Seng:2018qru} & \cite{Seng:2020wjq} & \cite{Czarnecki:2019mwq} & \cite{Hayen:2020cxh} & our estimate\\\midrule
	 Elastic & $1.05(4)$ & $1.06(6)$ & $1.06(6) $ & $0.99(10)$ & $1.06(6)$ & $1.06(6)$\\
	 Resonance & $0.04(1)$ & $0.05(1)$ & $0.05(1)$ & -- & -- & $0.04(1)$\\
	 Regge & $0.52(7)$ & $0.51(8)$ & $0.56(9)$ & $0.38^*$ & $0.46^*$ & $0.49(11)$\\\midrule
	 DIS & $2.29(3)$ & $2.26^*$ & $2.26^*$ & $2.24^*$ & $2.32^*$ & $2.28(4)$\\
\bottomrule
	\end{tabular}
	}
	\caption{Various contributions to $\Box_A^{\gamma W}$ in units of $10^{-3}$. The different columns are based on the references as indicated, but modified to correspond to the same conventions as far as possible (see main text), in particular, we use $Q_0^2=2\GeV^2$ in the separation of low- and high-energy parts and include the running of $\alpha$ in the evaluation of the DIS region (using the corrections from Ref.~\cite{Shiells:2020fqp} where necessary; modified entries are indicated by an asterisk and not assigned an uncertainty estimate). The DIS contribution enters for $Q^2\geq Q_0^2$ in Eq.~\eqref{master}, the rest for $Q^2\leq Q_0^2$. Note that the elastic contribution from Ref.~\cite{Czarnecki:2019mwq} is only integrated up to $1\GeV^2$, which explains the slightly smaller value. The resonance and Regge regions in Refs.~\cite{Seng:2018yzq,Seng:2018qru} are separated as indicated by Ref.~\cite{Seng:2020wjq}. For Refs.~\cite{Czarnecki:2019mwq,Hayen:2020cxh} the inelastic contributions for $Q^2\leq Q_0^2$ are booked in the ``Regge'' category.  This compilation is inspired by Table I in Ref.~\cite{Shiells:2020fqp}.}
	\label{tab:box}
\end{table}

For the $\gamma W$ box contributions we build upon the summary already compiled in Ref.~\cite{Shiells:2020fqp}, which leads to the estimates given in Table~\ref{tab:box}. Since most of the calculations use $Q_0^2= 2\GeV^2$, we use this value for the primary separation, and the comparison between $Q_0^2=1\GeV^2$ and $2\GeV^2$ from Ref.~\cite{Shiells:2020fqp} as an estimate for the translation. Similarly, we include the $4\%$ enhancement from keeping the $Q^2$ dependence of $\alpha(Q^2)$ in the evaluation of the DIS contribution as quoted in Ref.~\cite{Shiells:2020fqp}, which in Ref.~\cite{Czarnecki:2019mwq} enters  instead as an additional enhancement factor in Eq.~\eqref{master}. While there is reasonable agreement for the elastic and DIS regions once expressed in terms of the same conventions, some differences do remain in the Regge region of the integral, dominating the final uncertainty. To obtain a global estimate, we take in each category the naive average of the independent evaluations and assign an uncertainty that covers the spread in the results, as indicated in the last column in Table~\ref{tab:box}. The remaining uncertainties are then independent and can thus be added in quadrature, yielding our final estimates (based on Refs.~\cite{Seng:2018yzq,Seng:2018qru,Czarnecki:2019mwq,Seng:2020wjq,Hayen:2020cxh,Shiells:2020fqp})
\beq
\label{DeltaR}
\Delta_R^V=0.02467(27),\qquad 
\Delta_R=0.03983(27).
\eeq

\bibliographystyle{apsrev4-1_mod}
\balance
\biboptions{sort&compress}
\bibliography{Kmu}

\end{document}